\documentclass[12pt,a4paper]{article}
\usepackage{amsthm}
\theoremstyle{plain}
\usepackage{authblk}
\usepackage[english]{babel}
\usepackage{graphicx}
\usepackage{verbatim}
\usepackage{amsfonts}
\usepackage{amsmath}
\usepackage{txfonts}
\usepackage[T1]{fontenc}
\usepackage{color}
\usepackage{epsfig}
\usepackage{natbib}
\usepackage{url}
\usepackage[bookmarksnumbered=true, colorlinks=true, citecolor=blue]{hyperref}
\date{}

\title{A Metaphysical Reflection on the Notion of Background in Modern Spacetime Physics}
\author{Antonio Vassallo}
\affil{University of Lausanne, Department of Philosophy, CH-1015 Lausanne\\ \url{antonio.vassallo@unil.ch}}

\begin{document}

\maketitle
\begin{center}
Accepted for publication in L. Felline, F. Paoli, E. Rossanese (eds.) - \emph{New Developments in Logic and Philosophy of Science}.\\
\end{center}

\pdfbookmark[1]{Abstract}{abstract}
\begin{abstract}
The paper presents a metaphysical characterization of spatiotemporal backgrounds from a realist perspective. The conceptual analysis is based on a heuristic sketch that encompasses the common formal traits of the major spacetime theories, such as Newtonian mechanics and general relativity. It is shown how this framework can be interpreted in a fully realist fashion, and what is the role of background structures in such a picture. In the end it is argued that, although backgrounds are a source of metaphysical discomfort, still they make a spacetime theory easy to interpret. It is also suggested that this conclusion partially explains why the notion of background independence carries a lot of conceptual difficulties.\\
\\
\textbf{Keywords}: Background structure; spacetime theory; nomic necessity; dynamical sameness; principle of reciprocity; substantive general covariance; background independence.
\end{abstract}

\newcounter{contatore}
\setcounter{contatore}{0}
\newtheorem{fed}[contatore]{Definition}

\section{Introduction}
\begin{quote}
Tempus absolutum, verum, \& mathematicum, in se \& natura sua sine relatione ad externum quodvis, \ae quabiliter fluit [...] Spatium absolutum, natura sua sine relatione ad externum quodvis, semper manet similare \& immobile [...]\\
\citep[][p. 6]{418}
\end{quote}
Newtonian absolute space and time are the epitomes of background structures. Newton's definitions quoted above beautifully express the idea of a background spatiotemporal structure as something whose characteristic properties are insensitive to anything else. Such an idea is indeed straightforward but it is also a source of conceptual discomfort. Starting from the Leibniz/Clarke debate on Newtonian mechanics (NM), and continuing with the aether problem in classical electrodynamics, it became clearer and clearer that the assumption of absolute structures led to differences in the physical description that were not inherent in the phenomena.\\
These conceptual problems justified a ``war'' on Newtonian backgrounds that ended victoriously with general relativity (GR), which is quite uncontroversially considered the first spacetime theory that dispenses with background spatiotemporal structures - i.e., it is \emph{background independent}. However, despite the agreement over the fact that GR is a background independent theory, an uncontroversial definition of this feature is still missing. Having in mind the extremely intuitive characterization of background spatiotemporal structures in NM, we might frown upon this difficulty. The definition of a background independent theory seems straightforward: it is just a theory where no (spatiotemporal) structure bears its properties independently of anything else. Actually, things have proven much more difficult than this, as - for example - the discussion in \citet{47, 56} convincingly shows. The conceptual difficulties in spelling out what background independence exactly amounts to lead not only to interpretational problems for GR (think about the historical debate on the alleged ``generalized'' principle of relativity initially proposed by Einstein), but also makes it difficult to extend this framework to the quantum regime (see \citealp{435}, for a technically accessible introduction to the issue of background independence in quantum gravity).\\
The aim of this short essay is to contribute a reflection on the problem of background independence by revising the metaphysical characterization of spatiotemporal backgrounds under the light of modern spacetime physics. We will start by providing a heuristic sketch that highlights the formal traits that are common, at least, to the major spacetime theories such as NM, special relativity (SR), and GR. We will then discuss a possible way to interpret this unified framework in a straightforward manner, based on some minimal metaphysical commitments that will be assumed as working hypotheses. Finally, we will exploit this conceptual machinery to describe how a background structure would influence the physics of possible worlds where background dependent theories hold. The hope is that, from a  metaphysical analysis of possible worlds might come some hint to develop a better physical description of the actual one.

\section{A Primer on Spacetime Theories}
In order to simplify our metaphysical analysis, let us start by providing a simple formal sketch of a spacetime theory that is able to capture, albeit at a heuristic level, the theoretical traits that are common to the most important spacetime theories.\footnote{The following sketch is based on \citet{15}.} For simplicity's sake, we agree that a physical theory can be formalized as a set of relations between mathematical objects, and that each instantiation of such relations - once suitably interpreted - represents a possible state of affairs.\\
Our main concern, at this stage, is to propose a theoretically ductile picture of spacetime. The first step in this direction is to specify what the building blocks of spacetime are. Again, to keep things simple, we will just say that these primitive elements are called \emph{events}. After a theory is interpreted, then such elements will take a definite physical meaning, such as that of ``place-at-a-time'', or ``physical coincidence''. Claiming that spacetime is a set of events $\mathcal{M}$ is for sure general, but rather uninformative, which means that we need to add structure to it. The second step is, then, to equip the set of events with a notion of ``surroundings''. This can be achieved by defining a new set $M:=(\mathcal{M},\tau)$, which is nothing but our starting set $\mathcal{M}$ together with a family $\tau$ of its subsets satisfying the following requirements:
\begin{itemize}
\item [-] The empty set and $\mathcal{M}$ itself belong to $\tau$.
\item [-] Any union of arbitrarily many elements of $\tau$ is an element of $\tau$.
\item [-] Any intersection of finitely many elements of $\tau$ is an element of $\tau$.
\end{itemize}
$\tau$ is called a \emph{topology} on $\mathcal{M}$, and its elements are called \emph{open sets} in $M$. A subset $V$ of $M$ is a \emph{neighborhood} for an element $x\in \mathcal{M}$ iff there exists an open set $A\in\tau$ such that $x\in A\subseteq V$. Moreover, we require the elements of $\mathcal{M}$ to be topologically distinguishable and separable, i.e. for any two elements $x$ and $y$ of $\mathcal{M}$, there exists a neighborhood $U$ of $x$ and a neighborhood $V$ of $y$ such that their intersection is the empty set. In this way, we end up with a topological space $M$ with a well-defined criterion for judging whether any two events are numerically distinct or not.\\
The structure so defined over $M$ is sufficient to introduce a notion of continuity of a function, and this lets us apply a further constraint on the characterization of spacetime, that is, the fact that, locally, it has to appear Euclidean. This constraint is implemented by requiring that for any open set $A$ in $M$ there exist a function $h:A\rightarrow\mathbb{R}^{n}$  that is bijective, continuous and whose inverse is continuous. A function satisfying these conditions is called a \emph{homeomorphism}. Roughly speaking, this condition assures that, for any open set $A$ of $M$, all elements in $A$ can be labelled using a $n$-tuple of real numbers - which usually amounts to saying that $A$ admits a \emph{coordinatization} $\{x^{i}\}_{i=0,\dots,n-1}$. Furthermore, we want that, for each two coordinatizations on overlapping neighborhoods, the transition function from one coordinatization to the other - which is entirely defined and acting on $\mathbb{R}^{n}$ - is differentiable in the ordinary sense. If we have shaped our spacetime judiciously then, in general, to any coordinate transformation $\{x^{i}\}\rightarrow\{y^{i}\}$ defined in a neighborhood $A$ of $M$ corresponds a map $f:M\rightarrow M$ such that, for each point $P$ in $A$, $x^{i}(f(P))=y^{i}(P)$. It can be proven that such a map, also called \emph{intrinsic transformation}, preserves the structure defined so far on $M$. The set of all these structure-preserving transformations is nothing but the group $diff(M)$ of \emph{diffeomorphisms}\footnote{That is, those mappings from $M$ to itself which are bijective, continuous and differentiable together with their inverses.} acting on $M$. The reader not much fond of technicalities can just visualize $diff(M)$ as the group of permutations of elements of $M$ that represent smooth deformations of this space.\\
So far we have introduced some kind of ``canvas'' on which an even richer structure - consisting in a variety of geometrical objects - can be defined. The most simple example is that of a (continuous) curve, which is represented by a (continuous) map $\sigma:I\subseteq\mathbb{R}\rightarrow M$. In a given coordinate system  $\{x^{i}\}$, the curve acquires the form $x^{i}=x^{i}(t), t\in I$. Another possibility is to define a \emph{field-theoretic} object $\mathbf{\Theta}$ as a map from $M$ to another space $X$: if $X$ is a space of rank $2$ tensors, then $\mathbf{\Theta}$ will be a tensor field on $M$ whose components $\Theta_{ij}$ in a coordinate system  $\{x^{i}\}$ will be the elements of a $n\times n$ matrix. These geometrical objects can in general be transformed by the application of a diffeomorphism. For example, if we have a field $\Phi:M\rightarrow X$ and we want to apply to this field a transformation $f:M\rightarrow M$; this is done by defining such ``application'' as $f^{*}\Phi:=\Phi\circ f$ which, for all $x\in M$, means that $(f^{*}\Phi)(x)=\Phi(f(x))$. In case of a map $\gamma:I\rightarrow M$, instead, we have $f^{*}\gamma:=f\circ\gamma\Rightarrow(f^{*}\gamma)(y)=f(\gamma(y))$ for all $y\in\mathcal{Y}$. The fact that there is a (nearly) one-to-one correspondence between coordinate transition functions and diffeomorphisms allows us to switch from the coordinate language to the intrinsic one without caring for any loss of information.\\
Among all the geometrical objects definable over $M$, there is a subgroup of them that endow $M$ with more structure than just its topology - indeed, they supply $M$ with a \emph{geometry} properly said. The most important of these objects are the \emph{metric tensor} and the \emph{affine connection}. The former is a rank-$2$ tensor $\mathbf{g}$ that is symmetric (i.e. $g_{ij}=g_{ji}$ in all coordinate systems) and non-degenerate (i.e. the determinant $det|g_{ij}|$ of the matrix $|g_{ij}|$ is different from zero in all coordinate systems), and which makes it possible to define the notion of ``length'' of a curve on $M$. The latter is a derivative operator $\boldsymbol{\nabla}$ (also called \emph{covariant derivative}) that provides a precise meaning to the ``change of direction'' of a curve on $M$. Hence, for example, a curve that never changes direction is a \emph{straight line} or \emph{affine geodesic} on $M$. Since also $\mathbf{g}$ permits to define a straight line as the curve of shortest length between two points of $M$, we have also a notion of \emph{metric} geodesic which, in general, does not have to coincide with the affine one. For this reason, the connection is required to be compatible with the metric tensor, i.e. it must always be the case that $\boldsymbol{\nabla}\mathbf{g}=0$. Once we have a well-defined notion of straight line, we can tell ``how much'' it corresponds to the usual straight line of Euclidean geometry; this evaluation is made possible by the \emph{Riemann curvature tensor} $\mathbf{Riem}[\mathbf{g}]$. If the Riemann tensor is identically null all over the manifold, then the geodesics of $M$ are exactly those of Euclidean geometry, and we say that the spacetime is \emph{flat}, otherwise \emph{curved}.\\ 
Let us now make some concrete cases. The first example is perhaps the simplest one: the spacetime of special relativity (SR). This theory postulates a spacetime $M$ endowed with the Euclidean topology of $\mathbb{R}^{4}$, that is, there exists a homeomorphism mapping the entire manifold over $\mathbb{R}^{4}$. A metric tensor - the \emph{Minkowski} metric $\boldsymbol{\eta}$ - is defined over $M$. As expected this object takes the form of a $4\times 4$ matrix in whatever coordinate system. Moreover, it is always possible to find a coordinate system where $|\eta_{ij}|=diag(-1,1,1,1)$. The Minkowski metric is compatible with a flat connection that basically overlaps with the usual derivative operator of differential calculus: this means that, in SR, the geodesics of $M$ are the usual straight lines of Euclidean geometry.\\
In NM, things are more complicated. We still have that $M$ is globally homeomorphic to $\mathbb{R}^{4}$, but the geometric structure of the manifold is that of a bunch of Euclidean $3$-spaces piled together by a temporal $1$-flow - more compactly we write $M=E_{3}\times\mathbb{R}$. In order to achieve this structure, we need to postulate a Euclidean $3$-metric over each $3$-space plus a temporal metric that labels the succession of these spaces. We then fix a flat connection compatible with this building and, finally, we single out a particular class of straight lines that describes the trajectories of bodies at absolute rest. This class of geodesics fix a notion of ``sameness of place through time'', while the temporal metric evaluates time intervals in a coordinate-independent manner. In sum, this is the complicated machinery needed to depict an absolute space enduring over absolute time.\\
Finally, in the case of GR, there is no restriction either on the topology of $M$, or on the metric tensor $\mathbf{g}$, or on the affine connection $\boldsymbol{\nabla}$. The only conditions are that $\mathbf{g}$ and $\boldsymbol{\nabla}$ are compatible, and that $M$ is \emph{Lorentzian}, which means that it is always possible to find a coordinate system $\{x^{i}\}$ on a neighborhood $A$ of a point $P\in M$ such that \emph{exactly at that point} $\mathbf{g}$ reduces to the Minkowski metric.\\
In technical terms, all the spacetimes described above are instances of a $n$-dimensional (pseudo-)Riemannian manifold. In all cases we had $n=4$, but in general nothing prevents us from elaborating a theory where the manifold has higher dimensionality. In the Kaluza-Klein approach, for example, a further spatial dimension is added to spacetime, which hence is $5$-dimensional.\\
As we have seen from the above examples, the way we fix all the features of $M$, such as dimensionality, topology, geometry, or even further structures, varies from theory to theory. Some theories fix ab initio just few features, and let the others be dictated by the dynamics, while others presuppose from the outset rigid spatiotemporal structures that are not influenced by the dynamics. Obviously, these possible choices are relevant in determining whether a theory is background independent or not, as it will become clear later.\\
Now that we have given a formal account of spacetime, we are ready to define a spacetime theory in the following way:
\begin{fed}\label{st}
\textbf{(Spacetime theory)}
A \emph{spacetime theory} $\mathcal{T}$ is a set of mathematical relations $\mathfrak{E}$ involving a set of geometrical objects $\mathcal{O}$ defined over a $n$-dimensional Riemannian manifold $M$:
\end{fed}
\begin{equation}\label{uno}
\mathcal{T}=\mathcal{T}(M, \mathcal{O}; \mathfrak{E}).
\footnote{Just to be fair, it is not the case that a theory has to be formulated à la (\ref{uno}) in order to be considered a spacetime theory. There are, for example, cases of spacetime theories formulated in Lagragian terms, which cannot be cast in the form  (\ref{uno}). However, we do not have to mind this for the present purposes.}
\end{equation}
The power of (\ref{uno}) lies in the fact that this formal unification makes simpler to spell out the way a spacetime theory is usually interpreted. $M$ plus its additional geometrical structure is taken to be the spacetime properly called; a curve on $M$ describes the motion of a point-like particle (so it is called the \emph{worldline} of that particle), and a generic material field occupying a spacetime region $A$ is represented by a map which assigns to each point in $A$ a tensor (or a vector, or even a scalar). Hence, spacetime is ``decorated'' by particles' worldlines, which are more or less straight depending on the near presence of material fields, such as the electromagnetic one. If a field is able to bend the worldline of a particle and the particle is able to modify the configuration of a field, then the two are said to be \emph{interacting}.  All the possible interactions between physical objects and the resulting motions allowed on $M$ are expressed in terms of relations encoded in $\mathfrak{E}$, which, in a given coordinate system, take the form of differential equations involving the components of the geometrical objects. Here, as a working hypothesis, we will stick to this simple reading, which presupposes a realistic attitude towards the geometric objects of the theory. This means that we will consider all the geometric objects in $\mathcal{O}$ as referring either to real (or at least possible) objects or to properties born by them. Hence, for example, a curve on $M$ will commit us to the (possible) existence of point-like particles moving along that worldline. Since, in general, the objects in $\mathcal{O}$ are field-theoretic in nature, we will be also committed to the existence of fields, which, as we have seen, are further divided into geometric (e.g. metric tensor field) and material (e.g. the electromagnetic field). This ``doubly dualistic'' metaphysical stance involving mixed particle/field and geometry/matter commitments is of course naive and perfectible. However, the disagreeing reader can just take it as a mere choice of vocabulary, and still follow the conceptual analysis of background structures we are going to perform.\\
A key motivation to adopt a naive realist attitude towards $\mathcal{O}$ is that, by doing so, we have a more or less clear measure of how much structure a spacetime theory postulates. By claiming this, we accept the line of argument developed in \citet{436}, where it is argued that modern physical theories represent objective physical structures in terms of geometric field-theoretic objects. Hence, roughly speaking, the larger $\mathcal{O}$, the more structure is postulated by $\mathcal{T}$.\\
So far we have agreed to adopt, as a working hypothesis, a naively realistic attitude towards the geometrical objects $\mathcal{O}$ in (\ref{uno}), but this claim by itself is confusing: to what specific theory are we declaring our commitments? The answer is to \emph{all} the theories falling in the scope of definition \ref{st}, and this is our second working hypothesis. In order to better spell out this second assumption, we need to introduce another important definition:
\begin{fed}\label{model}
\textbf{(Model)}
A \emph{model} of a spacetime theory $\mathcal{T}$ is a $(k+1)$-tuple\\ $<M, \{ O_{k}\}_{k\in\mathbb{N}}>$ - where $O_{i}\in\mathcal{O}$ for all $i\leq k$ - that is a solution of $\mathfrak{E}$.
\end{fed}
If we think of the space $\mathcal{Q}_{\mathcal{T}}$ whose points represent each a configuration of \emph{all} the geometrical objects of the theory - which is in fact called \emph{configuration space} of the theory - then $\mathfrak{E}$ selects a subspace $\mathfrak{S}_{\mathcal{T}}\subset\mathcal{Q}_{\mathcal{T}}$ comprising all the physically allowed configurations of geometrical objects. This is at the root of the usual distinction between a purely kinematical state of affairs, that is, whatever element of $\mathcal{Q}_{\mathcal{T}}$, and a physical (or dynamical) state, which belongs to $\mathfrak{S}_{\mathcal{T}}$.\\
Definition \ref{model} concerns ``total'' or ``cosmological'' models, which means that, in a model $<M, \{ O_{n}\}>$, the geometrical objects are spread throughout the entire manifold $M$. However, it might be the case that a model admits a subclass of ``partial'' models involving a submanifold $\mathcal{K}\subset M$ and a set of geometrical objects defined on it.\\
 The concept of model is the most important one for interpretational purposes because, from a metaphysical point of view, a model of a theory represents a physically allowed state of affairs. According to our realist attitude, then, a cosmological model of a given theory $\mathcal{T}$ will represent an entire universe where the specific laws of $\mathcal{T}$ hold. In other words, it represents a nomically possible world. By the same token, a submodel of the same theory will be interpreted as a possible local states of affairs in a nomically possible world. In order to make the philosophical analysis easier, we will consider all and only the models of spacetime theories satisfying (\ref{uno}) and we will assume that this set of models represents a cluster of nomically possible situations. Each theory, then, individuates a subset of possible worlds where the particular laws $\mathfrak{E}$ of that theory are at work. Note that this working hypothesis does not restrict us to adopt a particular metaphysical stance neither with respect to possible worlds (they can be mental constructions as well as existent objects), nor with respect to laws of nature ($\mathfrak{E}$ can be either grounded, say, in some genuinely modal feature of the entities inhabiting a possible world, or can be just a description of regularity patterns crafted in that possible world).\\
The last important definition we need to put forward before digging into metaphysical considerations regards the notion of general covariance:
\begin{fed}\label{cov}
\textbf{(General covariance - Formal version)}
A spacetime theory $\mathcal{T}$ is \emph{generally covariant} iff, for all $f\in diff(M)$ and for all $\mathfrak{M}\in\mathfrak{S}_{\mathcal{T}}$, it is the case that $f(\mathfrak{M})\in\mathfrak{S}_{\mathcal{T}}$. $diff(M)$ is the \emph{covariance group} of $\mathcal{T}$.
\end{fed}
Here we talk of a ``formal'' version of general covariance  - as opposed to a ``substantive'' one, which we will encounter later - for the following reason. Since $\mathfrak{E}$ lives on the manifold $M$, i.e., it represents the way the geometrical objects of the theory are related throughout the manifold, and since $diff(M)$ is the group of the structure preserving mappings defined over $M$, then it is trivial to see that, by applying a diffeomorphism to whatever model of the theory, we obtain another model of the theory. Moreover, given that formal general covariance is trivially satisfied by any theory falling in the scope of definition \ref{st}, and given that it is possible to formulate extremely different physical theories in the form (\ref{uno}) - just think about the physical abyss that lies between NM and GR -, then it is clear that the notion of general covariance defined above is purely formal and bears no physical import (historically, \citealp{59}, was the first to acknowledge this fact).\\
A legitimate question might arise at this point. Given that radically different spacetime theories can be encompassed by the same formal framework, what is it exactly that makes them in fact radically different? To give a precise answer to this question, we need to say something more about the metaphysics of backgrounds.

\section{A Metaphysical Appraisal of Backgrounds}\label{mab}
The notion of background structure we are going to introduce draws from the work of \citet{65,66},\footnote{Further refined in \citet[][see in particular chapter II, sections 2 and 3]{15}.} and is based on the distinction made among the elements of $\mathcal{O}$ between dynamical and non-dynamical objects. Such a distinction will become clearer in a moment. For the time being, let us just say that a background structure $\boldsymbol{\mathcal{B}}\in\mathcal{O}$ is a geometrical object of the theory that is fixed ab initio and, hence, is ``persistent'' throughout the solution space of the theory.\\
To inform this notion with physics, consider the special relativistic description of the propagation of a massless scalar field:
\begin{equation}\label{dv}
\boldsymbol{\Box}_{\boldsymbol{\eta}}\boldsymbol{\phi}=0,
\end{equation}
where $\boldsymbol{\Box}_{\boldsymbol{\eta}}$ is the d'Alembertian operator with components $\eta_{ij}\frac{\partial}{\partial x^{i}}\frac{\partial}{\partial x^{j}}$ in some coordinate system.\footnote{The Einstein convention is applied here.} Let us further assume that (\ref{dv}) has two solutions $\boldsymbol{\phi}_{1}$ and $\boldsymbol{\phi}_{2}$. According to our metaphysical hypotheses, this means that the SR-cluster admits two possible worlds that are described by the models $<\boldsymbol{\eta},\boldsymbol{\phi}_{1}>$ and $<\boldsymbol{\eta},\boldsymbol{\phi}_{2}>$. It is obvious to claim that these two worlds share a single feature, namely, the Minkowski metric. The key point is that we can repeat this operation with any two special relativistic worlds, that is, if we inspect the entire space of models of SR, we see that \emph{all} the models of the theory feature $\boldsymbol{\eta}$. From our metaphysical perspective, this translates to the fact that, in all possible worlds belonging to the SR-cluster, there always exists a Minkowski spacetime. Generalizing, we can think of characterizing a background structure by means of its metaphysical necessity or, better, its nomic necessity: a background structure $\boldsymbol{\mathcal{B}}$ of a given spacetime theory $\mathcal{T}$ is an object that such a theory deems necessary, i.e., there are no possible worlds described by $\mathcal{T}$ where $\boldsymbol{\mathcal{B}}$ does not exist.\\
Along with this first metaphysical feature of background structures comes a clear reason to feel uncomfortable with background dependent theories. A theory that postulates a necessary physical structure is conceptually puzzling, not least because it tells us that there is just \emph{one} physical possibility among many conceivable ones. By the same token, taking a structure as nomically necessary entails that it is physically impossible for it to change although we can conceive of a process in which the structure under scrutiny might in fact change. From an epistemic perspective, we can say that, when a theory accords a nomically necessary status to a spatiotemporal structure $\mathcal{B}$, then it is unable to provide a physically justified answer to the question ``why is it $\mathcal{B}$ and not otherwise?''. In the case of SR, the theory tells us that the only physically possible spacetime is the Minkowski one, and the only answer this theory can provide to the question ``why is it not otherwise?'' is ``because it is how it is''. Some may object that there is nothing really conceptually puzzling here, since it is totally reasonable to expect that the chain of physical justifications provided by a theory stops somewhere - i.e. there always comes a point in which a theory can just answer ``because it is how it is''. This is fair enough. However, this does not prevent us from putting two claims on the table. The first is: the fewer objects in $\mathcal{O}$ a theory deems nomically necessary, the better. This is because, then, such a theory is likely to exhibit a deeper explanatory structure than other spacetime theories that are more metaphysically ``rigid''. For example, GR is better than SR with this regard because it explains why and under what circumstances spacetime has a Minkowskian structure. Of course, this claim is not sacrosanct, in the sense that surely some counter-examples can be mounted against it. However, it still is fairly reasonable if applied to the major spacetime theories we have so far. The second claim we want to highlight is: it is not impossible that a theory falling in the scope of definition \ref{st} does not commit us to the nomic necessity of any of the objects in $\mathcal{O}$. Clearly, this second claim does not entail that such a theory admits a bottomless structure of physical justification - although many philosophers would not find anything wrong with that -, but just that the theory fixes ab initio some features other than (full) spatiotemporal structures.\\
A second important metaphysical feature of spatiotemporal backgrounds comes from the following example. Let us focus on the Newtonian cluster of possible worlds, and consider a Newtonian world where there exist a large ship docked on a calm sea. Inside the ship, shut up in the main cabin below decks, there is a man - we can call him Salviati - together with an experimental equipment consisting of jars of flies, fishes in bowls and dripping bottles.\footnote{Here, of course, we are referring to the ``Gran Naviglio'' thought experiment in \citet{319}.}
Simply speaking, we are dealing with a global model $\mathfrak{M}$, which describes the possible world in its entirety, but we are magnifying just a portion of it, that is a submodel $\mathfrak{m}$ describing just what happens in the immediate surroundings of the ship. Let us now apply to $\mathfrak{m}$ a transformation $f$ that consists in a rigid spatial translation of the ship. The model $f^{*}\mathfrak{m}$ will then depict a situation in which the ship is still on a calm sea without wind, but now it is located, say, one meter away from the position it had in $\mathfrak{m}$. In what dynamical aspects does $\mathfrak{m}$ and $f^{*}\mathfrak{m}$ differ? None: in both cases the ship is at absolute rest and Salviati is unable to spot any difference by looking at the equipment on board. This reasoning can be repeated with rotations. Take $f$ as a $45°$ rotation of the ship with respect to the original orientation, and again both $\mathfrak{m}$ and $f^{*}\mathfrak{m}$ will depict a ship at absolute rest, where Salviati's equipment behaves exactly in the same manner as the non-rotated one. We then suspect that the notion of sameness for Newtonian states of affairs is influenced by the underlying background structures. In this case, since Euclidean space is homogeneous and isotropic, the state of absolute rest of the ship is insensitive to where the ship is placed or how it is oriented.\\ As an acid test, consider another situation where $f^{*}\mathfrak{m}$ makes Salviati's ship sailing over troubled waters. In this case, it is quite obvious that $\mathfrak{m}$ and $f^{*}\mathfrak{m}$ depict radically different dynamical situations. The ship in $f^{*}\mathfrak{m}$ is not in a state of absolute rest (its worldline is not a geodesic at all, let alone a straight line pointing in the privileged ``rest direction''), and this has quite disruptive observable consequences: while in $\mathfrak{m}$ Salviati sits down quietly observing his jars of flies, fishes in bowls and dripping bottles, in $f^{*}\mathfrak{m}$ he\footnote{Or, if you want, his counterpart, depending on the particular account of possible worlds adopted.} 
is shaking in the main cabin among broken glasses, buzzing flies and asphyxiating fishes.\\
To sum up, we have individuated another very important metaphysical aspect of backgrounds, namely, that they fix a notion of sameness of dynamical state throughout the cluster of possible worlds of the theories they figure into. From a formal perspective, this means that, if a spacetime theory admits a set of background structures $\{\boldsymbol{\mathcal{B}}_{i}\}$, then for whatever two models of the theory related by some transformation $f$, these two models are said to be dynamically indiscernible iff $f^{*}\boldsymbol{\mathcal{B}}_{i}=\boldsymbol{\mathcal{B}}_{i}$, for all $i$, that is, iff $f$ is a transformation (called \emph{isometry}) that leaves all the background structures invariant. We call this set of isometries $iso(\{\boldsymbol{\mathcal{B}}_{i}\})\subset diff(M)$ the \emph{symmetry group} of the theory.\\
This definition of symmetry qualifies as ``ontic'' in the taxonomy put forward in \citet{437}. The author charges this kind of definition with inferential circularity. In his own words:
\begin{quote}
But according to an ontic definition of `symmetry', in order to check whether a given transformation [$f$] counts as a symmetry of [dynamical] laws, I first need to know which physical features fix the data so that I can check whether [$f$] preserves them. And the problem is that, in many cases, we discover which physical features fix the data by engaging in
symmetry-to-reality reasoning!\\
(\emph{Ibid, p. 28})
\end{quote}
Although the above issue is a serious one, worth of extensive philosophical discussion, here we just dodge the charge of inferential circularity by appealing to our naive realist framework. Simply speaking, we do not discover which physical features ``fix the data'' (in our case, the background structures): we just postulate them ab initio.\\
At this point, we can go back to the question raised at the end of the previous section, that is, what is it that renders different spacetime theories in fact different? The answer is now crystal clear: the background structures in $\mathcal{O}$. It is in fact thanks to the backgrounds postulated by a theory that we can attribute physical import to a subset of the covariance group $diff(M)$. We have then different theories depending on the subset individuated by the backgrounds. For example, we can say that NM is physically different from SR because the former admits a set of symmetries which form a group called \emph{Galilean}, while the symmetries of the latter belong to the \emph{Poincar\'e} group.\\
However, as in the previous case, also this feature of backgrounds may lead to unhappy consequences. To see this let us consider again the docked ship on a calm sea in $\mathfrak{m}$, and transform this model in one where the ship is still on a calm sea, but now it is sailing with uniform velocity. Technically, the transformation $f$ involved in this case belongs to the so-called Galilean group. Intuitively speaking, while in $\mathfrak{m}$ the ship is in a trajectory of absolute rest (straight line pointing in the privileged direction), $f$ just ``inclines'' the trajectory of an arbitrary angle without ``bending'' it. We are now in a strange situation: from the global perspective of $\mathfrak{M}$, $\mathfrak{m}$ and $f^{*}\mathfrak{m}$ depict different dynamical states - absolute rest vs. motion with uniform absolute velocity, but from Salviati's perspective, there is no empirically observable difference between the two dynamical states! Here, as in the case of nomic necessity, a liberal metaphysician might claim that we should not worry too much and just accept the fact that our theory commits us to the existence of dynamically different yet empirically indistinguishable states of affairs. After all, this is just a metaphysical fact that does not impair in any way the role of physicists. In fact, it is obvious that whatever empirical  question regarding the dynamics that Salviati could ask would \emph{always} have an answer, which would be the same irrespective of the fact that the ship is in a state of absolute rest or absolute uniform motion. Again, we concede the point that the existence of dynamically distinct yet empirically indistinguishable states of affairs is not a mortal sin for a theory. But accepting this means accepting that there can be elements of reality that are \emph{totally} opaque to physics! This is a rather embarrassing claim to embrace, especially if we believe that metaphysics must be motivated and informed by science (and physics in particular). At least, it is reasonable to invoke some sort of Occamist norm according to which, among two competing theories with the same empirical consequences, we should prefer the one that commits us to the least structure. Let us try to apply such a norm to NM.\\
The evidence that the culprit for the above discussed unwanted situation is absolute space is given by the fact that the Galilean group is part of the isometries of all Newtonian background objects except for the class of straight lines that fixes the notion of ``sameness of place through time''. Fortunately, we can reformulate NM without privileging any set of geodesics and, hence, giving up the commitment to absolute space.\footnote{As shown, for example, in \citet[][Chapter III, section 2]{15}.} In this new framework this particular problem evaporates since now the dynamics of the theory does not distinguish anymore states of rest from states of uniform velocity.\\
In sum, here lies the second charge against background structures: the more background structures a theory admits, the more it is likely that the theory will consider as dynamically distinct some models that, in fact, admit the very same physical observables.\\
The last metaphysical feature of a background structure is related to the distinction between dynamical and non-dynamical objects mentioned at the beginning of the section. In short, spatiotemporal backgrounds are non-dynamical objects because they do not enter $\mathfrak{E}$ as elements subjected to the dynamical laws but, rather, they represent the support that renders possible the very formulation of such laws. The problem with the non-dynamicity of background structures is summarized in the following quote:
\begin{quote}
[A]n absolute element in a theory indicates a lack of reciprocity; it can influence the physical behavior of the system but cannot, in turn, be influenced by this behavior. This lack of reciprocity seems to be fundamentally unreasonable and unsatisfactory. We may express the converse in what might be called a general principle of reciprocity: Each element of a physical theory is influenced by every other element. In accordance with this principle, a satisfactory theory should have no absolute elements.\\
\citep[][p. 192]{65}
\end{quote}
Anderson effectively summarizes the third peculiarity of backgrounds and the reason why we should feel uneasy about that. However, few comments are in place. First of all, the way Anderson enunciates the principle of reciprocity is too strong and seems to amount to some holistic principle which, most likely, was not the author's intention. Perhaps it would have been better to say that each element of a physical theory \emph{can be} influenced by \emph{some} other element. Secondly, the principle as it stands can be easily challenged on the ground of its vagueness as to how an ``element of a physical theory'' has to be understood. To see why it is so, we could just consider the Lagrangian formulation of NM. In this framework, the behavior of a mechanical system is fully described by the Lagrange equations: once we fix an appropriate Lagrangian plus initial conditions, we get the full dynamical history of the system in the form of a trajectory in configuration space. In a sense, then, the Lagrangian function is an element of the theory that influences  the mechanical system but that is not influenced back, being it a supporting element of the dynamical description. Does it imply that the Lagrangian violates the principle of reciprocity? Here, we are exploiting the vagueness underlying the notion of ``element of a physical theory''. The Lagrangian is with no doubt an element of the theory, but it would be awkward to interpret it as ontologically on a par with the mechanical system: it is just a descriptive tool that carries dynamical information and, as such, has not to be taken as referring to a concrete object that exists over and above the mechanical system. Evidently, a too broad characterization of an element of the theory led us to a category mistake.\\
Fortunately, the theoretical framework given by definitions \ref{st} and \ref{model} helps us clarifying the real intentions behind Anderson's quote above. If, in fact, we restrict the scope of the principle of reciprocity to the geometrical objects definable over $M$, we can restate the principle as follows: each element of the set $\mathcal{O}$ must be subjected to the dynamical evolution encoded by $\mathfrak{E}$. This renders the principle of reciprocity less vague and highlights in what sense Anderson characterizes background structures as elements of the theory that violate such a principle. However, we still have the possibility to scupper this characterization. To do so, it is sufficient to reconsider the example of the theory with equation (\ref{dv}). As we have seen, this theory features a background structure, namely the Minkowski metric $\boldsymbol{\eta}$. Now, let us add to (\ref{dv}) a further equation:
\begin{equation}\label{dv2}
\mathbf{Riem}[\mathbf{g}]=0.
\end{equation}
What have we done here? Leaving aside technical considerations, we have done nothing but ``embedding'' the fixing condition of the Minkowski metric into $\mathfrak{E}$. Hence, the solution space of this new theory carries absolutely no more physical information than the one associated to (\ref{dv}) alone, and the Minkowski metric is still a background structure satisfying the first two features we have reported. However, now, we have a theory that challenges the utility of the principle of reciprocity as a guide in assessing spacetime theories. In the theory (\ref{dv})/(\ref{dv2}) each element of the set $\mathcal{O}$ is subjected to the dynamical evolution encoded by $\mathfrak{E}$, but still the theory admits a background. This example shows that even the amended version of the principle of reciprocity we have considered is conceptually flawed. Nonetheless, it seems still evident that Anderson's quotation captures a salient feature of backgrounds. Perhaps, we should read this quote in a more straightforward way, and interpret the talk in term of influences as referring to a very concrete notion of physical interaction. In some sense, here we are shifting the problem to what exactly ``interacting'' amounts to in the modern physical jargon. However, just for the sake of argument, let us assume that an interaction between two elements $\mathbf{\Theta}_{1}$ and $\mathbf{\Theta}_{2}$ of a theory amounts to adding to $\mathfrak{E}$ a coupling relation of the form $F(\mathbf{\Theta}_{1},\mathbf{\Theta}_{2},\kappa)$, $\kappa$ being and appropriate coupling parameter. If we reconsider the principle of reciprocity under this light, than it becomes the statement that each field-theoretic object is coupled with some other. The challenge of the theory (\ref{dv})/(\ref{dv2}) is now defused because the background role of the Minkowski metric is restored due to the fact that it does not satisfy this latter version of the principle of reciprocity. Therefore, in the end, we can say that the third metaphysical feature of spatiotemporal background is the one already highlighted by Newton's quotation at the beginning of the paper, namely that they bear their properties without relation to anything else: this feature can be reasonably translated in the language of modern spacetime physics as the fact that they are structures that are not coupled to any material field.\\
Is this a bad thing, metaphysically speaking? Let us answer with the words of Brown and Lehmkuhl:
\begin{quote}
If there is a questionable aspect of [the principle of reciprocity], it is less the claim that substances act (how otherwise could their existence be known to us?) than the notion that they are necessarily acted back upon, that action must be reciprocal. If all substances act, they do so in relation to other substances; these other substances therefore cannot be immune from external influences. Now it might seem arbitrary on \emph{a priori} grounds to imagine that the `sensitivity' of such substances is not universal. That is to say, it might seem arbitrary to suppose that not all substances react to others. But no such abstract qualms can be entirely compelling; Nature must have the last say.\\
\citep[][pp. 3, 4]{438}
\end{quote}
Otherwise said, pursuing the principle of reciprocity is reasonable but not necessary. To further reflect on this point, let us focus on NM and ask in what sense the absolute backgrounds of this theory influence the motion of bodies. For example, what is it that ``forces'' an isolated point-particle to move in a straight line? The answer is obviously ``nothing'', let alone absolute structures: it is just a primitive fact - i.e. non further justifiable via a ``why'' question - that in every Newtonian world there exists a privileged class of trajectories occupied by bodies in inertial motion. In this sense, absolute structures \emph{define} possible motions but do not push (in a ordinary physical sense) bodies to move that way. Under this light, it does not seems that conceptually hard to withstand a violation of the principle of reciprocity.

\section{Conclusion: How Easily Can We Dispense with Backgrounds?}
In the previous section we have supplied a metaphysical characterization of spatiotemporal backgrounds based on the language of modern spacetime physics. To recap, we have highlighted three features of background structures in a spacetime theory:
\begin{enumerate}
\item The theory in which they feature treats them as (nomically) necessary structures.
\item They induce a notion of dynamical sameness among states of affairs throughout the solution space of the theory.
\item Their dynamical influences are not describable as physical interactions.
\end{enumerate}
As we have discussed, with each of this metaphysical traits comes an associated conceptual discomfort. However, we have also highlighted that none of these issues lead to contradictions or physical loopholes. Hence, we are inclined to claim that whether one wants to renounce background structures depends on one's own metaphysical tastes. Otherwise said, one can backup one's commitment to background independence with strong and convincing arguments (and, indeed, many of such arguments can be found in the literature), but she cannot appeal to a requirement of background independence as a physically necessary one.\\
However, the realist framework we have put forward has made clear that background structures have not only (mild) metaphysical vices, but also metaphysical virtues. The most important among them is the possibility to straightforwardly define the notion of physical symmetry in an ontic manner, without incurring inferential circularity. More generally, once we specify what are the background structures $\{\boldsymbol{\mathcal{B}}_{i}\}$ of a theory, the interpretation of such a theory becomes a rather smooth business: this is because, once the symmetries of a theory are given, we can identify as referring to real objects or properties those theoretical structures that are invariant under these symmetries. Once again, we stress that this is possible because we assume background structures as postulated ab initio as a matter of ontological fact. In general, in fact, there is no formal criterion that makes an object in $\mathcal{O}$ a background structure, and it can be the case that the very same geometric object can count or not count as background depending on the particular interpretation of the theory chosen \citep[][section 3.3, discusses in detail the case of such geometrically ambiguous theories]{311}.\\
So far we have engaged in a conceptual cost-benefit analysis of postulating background structures in our theory. Suppose, now, that we are inclined to buy into the view that a background has more costs than benefits and, hence, we wish to go for background independence. According to our framework, implementing such a requirement amounts - at least - to constructing a theory whose spatiotemporal structures do not satisfy the three conditions listed at the beginning of the section. Here, obviously, we cannot undertake this task, so we will be just content to verify whether GR, which is usually considered the epitome of background independent theory, in fact violates the three metaphysical requirements for background structures.\\
The dynamical equations of GR have the form $\mathbf{G}[\mathbf{g}]=\kappa\mathbf{T}[\boldsymbol{\phi},\mathbf{g}]$, where the left-hand side of the relation represents the geometry of spacetime (the so-called Einstein tensor), and the right hand side features the stress-energy tensor, which encodes information regarding the mass-energy distribution over a region of spacetime. We can then say that spacetime in GR is not a background in primis because the theory is about the coupling of the metric field $\mathbf{g}$ with the matter field(s) $\boldsymbol{\phi}$ and, hence, the third requirement above is not met. From the form of the dynamical equations, in the second place, we infer that it is not the case that all the models of the theory feature the same geometric objects and, hence, in the GR-cluster of possible worlds there is no field-theoretic structure that counts as nomically necessary.\footnote{As a matter of fact, some examples might be provided, which challenge this claim (see, e.g., \citealp{354}). However, since these examples are not disruptive to our analysis, we can set them aside.} It seems, then, that also the second condition is not fulfilled. As a matter of fact, as we have hinted at in section \ref{mab}, there are other features of the models of the theory that bear a physical significance and that show the ``persistence'' typical of backgrounds. For example, all models of GR feature manifolds of dimensionality $4$ and Lorentzian in nature. Hence, although GR does not treat any spatiotemporal structure as nomically necessary, there are some characteristic traits of these structures that are nonetheless preserved throughout the solution space of the theory.  Hence, strictly speaking, in GR the spatiotemporal structures do bear at least some properties without relation to anything external.\\
To get rid once and for all of this kind of objections, we can somehow render our distinction between background dependent and independent theories more flexible. Up to now, in fact, we have assumed that, in order to consider a theory background dependent, it is sufficient that it admits at least a background structure. However, this sort of classification might be too coarse
or might deliver an unintuitive picture. Consider for example a theory whose equations have two classes of models: one featuring, say, a flat metric, and another featuring a curved one. Clearly, these two metrics would not qualify as backgrounds according to the above characterization, since they are not nomically necessary objects according to the theory. Still, we would feel unconfortable with this conclusion, since such a theory would still be ``ontologically rigid''. Perhaps, we can establish a well-defined way to count (i) how many physical features in general - not only geometric objects in $\mathcal{O}$ - are deemed nomically necessary by the theory and (ii) how often non-nomically necessary features appear throughout the solution space of the theory. This would imply that the distinction between background dependence and independence would not be so clear-cut, there being different degrees in which they come. If this strategy can be consistently worked out (\citealp{311}, makes a concrete proposal along these lines), then we would have a measure according to which, say, NM is fully background dependent, while GR is fully background independent modulo minor fixed features.\\
Finally, let us consider the second requirement and ask, if GR has no background structures, does it still possess a well-behaved notion of dynamical sameness? We face a dilemma here: if we answer no, this would imply that GR is a useless theory incapable of making even the simplest empirical predictions, which is most obviously not the case; if we answer yes, then we have to face a huge controversy. To see why it is so, let us back 
up our affirmative answer with the following argument:
\begin{itemize}
\item [(P1)] The physical symmetries of a spacetime theory are those transformations $f\in diff(M)$ that are isometries for the background structures $\{\boldsymbol{\mathcal{B}}_{i}\}$;
\item [(P2)] GR has no background structures, i.e. $\{\boldsymbol{\mathcal{B}}_{i}\}=\emptyset$;
\end{itemize}
Therefore,
\begin{itemize}
\item [(C)] In GR, \emph{all} transformations $f\in diff(M)$ are physical symmetries of the theory.
\end{itemize}
The conclusion of this argument is usually stated as the fact that GR satisfies the requirement of \emph{substantive} general covariance, as opposed to the mere formal version given by definition \ref{cov}. Note that a similar argument can be mounted, in which (P2) and (C) are switched. In this way, background independence and substantive general covariance would become overlapping concepts. The problem with this line of argument is that it forces us to buy into the view that, trivially, the transformations in $diff(M)$ are all at once isometries of \emph{no} background structure (whatever diffeomorphism applied to nothing does not change anything). But that seems too loose an appeal because the distinction between the whole $diff(M)$ and $iso(\{\boldsymbol{\mathcal{B}}_{i}\})$ \emph{requires} background structures: if such structures are absent, then we have no means for making the distinction. By the same token, starting from the premise that all diffeomorphisms are physical symmetries of the theory does not provide a firm enough ground to infer that the theory is background independent, since we can always disguise background structures as dynamical objects.\\
Hence, it seems clear that, in order to define substantive general covariance in a more rigorous way, it is necessary to base the argument for having $diff(M)$ as the set of physical symmetries on an approach different from the one considered in this paper. \citet{174}, for example, analyzes substantive general covariance in terms of variational symmetries in the Lagrangian formalism, but this approach does not help with spacetime theories that cannot be rendered in Lagragian terms.\footnote{See \citet{370} for a detailed criticism of Earman's proposal.}
\citet{391}, instead, argues that the problem arises from a wrong way of looking at the structure of spacetime theories. Very simply speaking, Stachel claims that the physically relevant information regarding a spacetime theory is not in general encoded in the manifold $M$, but in a more complex structure, namely, a triple of topological spaces - technically called fiber bundle - $(\mathfrak{X}, M,\mathcal{F})$, with $\mathfrak{X}$ having locally the form $M\times\mathcal{F}$. In this context, the dynamical equations $\mathfrak{E}$ become a set of rules for selecting cross-sections of this fiber bundle.\footnote{Intuitively, if the fiber bundle is a simple vector bundle, then a cross-section of it would be a vector field over $M$.} Now, the requirement of substantive general covariance amounts to the fact that all the (geometrical objects referring to) spatiotemporal structures of the theory live on these cross-sections. If some structure still lives on the manifold $M$, then the theory is background dependent.\\
Stachel's approach might prove more effective than that represented by (\ref{uno}) in highlighting the formal differences between spacetime theories - especially with respect to considerations regarding background dependence/independence. However, it does not seem to bring much ontological clarity to the matter. While, in fact, the framework we put forward admits a straightforward interpretation, it is not at all clear how to spell out the way the structure $(\mathfrak{X},M,\mathcal{F})$ refers to real (or possible) physical structures.\\
In conclusion, the most important moral we can draw from the analysis developed in this paper is that background structures, albeit showing some metaphysical vices, are nonetheless elements that render the formulation and the interpretation of a spacetime theory sharp and fairly simple. This is why pursuing the requirement of background independence demands a huge conceptual price to be paid.

\pdfbookmark[1]{Acknowledgements}{acknowledgements}
\begin{center}
\textbf{Acknowledgements}:
\end{center}
I wish to thank an anonymous referee and Davide Romano for helpful comments on an earlier version of this paper. Research contributing to this paper was funded by the Swiss National Science Foundation, Grant no. $105212\_149650$.

\pdfbookmark[1]{References}{references}
\bibliography{biblio}

\begin{thebibliography}{}

\bibitem[\protect\citeauthoryear{Anderson}{Anderson}{1964}]{65}
Anderson, J. (1964).
\newblock Relativity principles and the role of coordinates in physics.
\newblock In H.~Chiu and W.~Hoffmann (Eds.), {\em Gravitation and Relativity},
  pp.\  175--194. W.A. Benjamin, Inc.

\bibitem[\protect\citeauthoryear{Anderson}{Anderson}{1967}]{66}
Anderson, J. (1967).
\newblock {\em Principles of relativity physics}.
\newblock Academic Press.

\bibitem[\protect\citeauthoryear{Belot}{Belot}{2011}]{311}
Belot, G. (2011).
\newblock Background-independence.
\newblock {\em General Relativity and Gravitation\/}~{\em 43}, 2865--2884.
\newblock \url{http://arxiv.org/abs/1106.0920}.

\bibitem[\protect\citeauthoryear{Brown and Lehmkuhl}{Brown and
  Lehmkuhl}{2015}]{438}
Brown, H. and D.~Lehmkuhl (2015).
\newblock Einstein, the reality of space, and the action-reaction principle.
\newblock Forthcoming in Ghose, P. (ed.) \emph{The nature of reality}.
  \url{http://arxiv.org/abs/1306.4902}.

\bibitem[\protect\citeauthoryear{Dasgupta}{Dasgupta}{2015}]{437}
Dasgupta, S. (2015).
\newblock Symmetry as an epistemic notion (twice over).
\newblock {\em British Journal for the Philosophy of Science\/}.
\newblock {DOI} 10.1093/bjps/axu049.

\bibitem[\protect\citeauthoryear{Earman}{Earman}{2006}]{174}
Earman, J. (2006).
\newblock Two challenges to the requirement of substantive general covariance.
\newblock {\em Synthese\/}~{\em 148\/}(2), 443--468.

\bibitem[\protect\citeauthoryear{Friedman}{Friedman}{1983}]{15}
Friedman, M. (1983).
\newblock {\em Foundations of Space-Time Theories. {R}elativistic Physics and
  Philosophy of Science}.
\newblock Princeton University Press.

\bibitem[\protect\citeauthoryear{Galilei}{Galilei}{1632}]{319}
Galilei, G. (1632).
\newblock {\em Dialogo sopra i due massimi sistemi del mondo tolemaico e
  copernicano}.
\newblock Giovanni Battista Landini.

\bibitem[\protect\citeauthoryear{Giulini}{Giulini}{2007}]{47}
Giulini, D. (2007).
\newblock Remarks on the notions of general covariance and background
  independence.
\newblock {\em Lecture notes in physics\/}~{\em 721}, 105--120.
\newblock \url{http://arxiv.org/abs/gr-qc/0603087}.

\bibitem[\protect\citeauthoryear{Kretschmann}{Kretschmann}{1917}]{59}
Kretschmann, E. (1917).
\newblock {\"U}ber den physikalischen sinn der relativit\"atspostulate, {A}.
  {E}insteins neue und seine urspr\"ungliche {R}elativit\"atstheorie.
\newblock {\em Annalen der Physik\/}~{\em 53}, 575--614.
\newblock Italian translation by S. Antoci available at
  \url{http://fisica.unipv.it/antoci/re/Kretschmann17.pdf}.

\bibitem[\protect\citeauthoryear{Newton}{Newton}{1726}]{418}
Newton, I. (1726).
\newblock {\em Philosophiae Naturalis Principia Mathematica\/} (Third ed.).
\newblock The Royal Society of London.

\bibitem[\protect\citeauthoryear{North}{North}{2009}]{436}
North, J. (2009).
\newblock The "structure" of physics: {A} case study.
\newblock {\em Journal of Philosophy\/}~{\em 106}, 57--88.
\newblock \url{http://philsci-archive.pitt.edu/4961/}.

\bibitem[\protect\citeauthoryear{Pitts}{Pitts}{2006}]{354}
Pitts, J. (2006).
\newblock Absolute objects and counterexamples: {J}ones-{G}eroch dust,
  {T}orretti constant curvature, tetrad-spinor, and scalar density.
\newblock {\em Studies in History and Philosophy of Modern Physics\/}~{\em
  37\/}(2), 347--351.
\newblock \url{http://arxiv.org/abs/gr-qc/0506102v4}.

\bibitem[\protect\citeauthoryear{Pooley}{Pooley}{2010}]{370}
Pooley, O. (2010).
\newblock Substantive general covariance: {A}nother decade of dispute.
\newblock In M.~Su\`arez, M.~Dorato, and M.~R\`edei (Eds.), {\em EPSA
  Philosophical Issues in the Sciences: {L}aunch of the European Philosophy of
  Science Association}, Volume~2, Chapter~19, pp.\  197--209. Springer.
\newblock \url{http://philsci-archive.pitt.edu/9056/1/subgencov.pdf}.

\bibitem[\protect\citeauthoryear{Rickles}{Rickles}{2008}]{56}
Rickles, D. (2008).
\newblock Who's afraid of background independence?
\newblock In D.~Dieks (Ed.), {\em The ontology of spacetime}, Volume~2 of {\em
  Philosophy and foundations of physics}, Chapter~7, pp.\  133--152. Elsevier
  B.V.
\newblock \url{http://philsci-archive.pitt.edu/4223/}.

\bibitem[\protect\citeauthoryear{Rozali}{Rozali}{2009}]{435}
Rozali, M. (2009).
\newblock Comments on background independence and gauge redundancies.
\newblock {\em Advanced science letters\/}~{\em 2\/}(3), 244--250.
\newblock \url{http://arxiv.org/abs/0809.3962v2}.

\bibitem[\protect\citeauthoryear{Stachel}{Stachel}{1986}]{391}
Stachel, J. (1986).
\newblock What a physicist can learn from the discovery of general relativity.
\newblock In R.~Ruffini (Ed.), {\em Proceedings of the {F}ourth {M}arcel
  {G}rossmann meeting on general relativity}, pp.\  1857--1862. Elsevier B.V.

\end{thebibliography}
\end{document}